# Application of Machine Learning Optimization in Cloud Computing Resource Scheduling and Management

Yifan Zhang [1]*
Executive Master of Business Administration
The University of Texas at Arlington
Arlington, Texas, USA
* Corresponding author:yifan.ibm@gmail.com

Bo Liu [2]
Software Engineering
Zhejiang University
HangZhou China
lubyliu45@gmail.com

Yulu Gong [3]
Computer & Information Technology
Northern Arizona University
Flagstaff, AZ, USA
yg486@nau.edu

Jiaxin Huang [4]
Information Studies
Trine University
Phoenix USA
jiaxinhuang1013@gmail.com

Jingyu Xu [5]
Computer Information Technology
Northern Arizona University
Flagstaff, AZ,USA
jyxu01@outlook.com

Weixiang Wan [6]
Electronics & Communication Engineering
University of Electronic Science and Technology of China
Chengdu, China
danielwanwx@gmail.com

*Abstract*— **In recent years, cloud computing has been widely used. Cloud computing refers to the centralized computing resources, users through the access to the centralized resources to complete the calculation, the cloud computing center will return the results of the program processing to the user. Cloud computing is not only for individual users, but also for enterprise users. By purchasing a cloud server, users do not have to buy a large number of computers, saving computing costs. According to a report by China Economic News Network, the scale of cloud computing in China has reached 209.1 billion yuan. At present, the more mature cloud service providers in China are Ali Cloud, Baidu Cloud, Huawei Cloud and so on. Therefore, this paper proposes an innovative approach to solve complex problems in cloud computing resource scheduling and management using machine learning optimization techniques. Through in-depth study of challenges such as low resource utilization and unbalanced load in the cloud environment, this study proposes a comprehensive solution, including optimization methods such as deep learning and genetic algorithm, to improve system performance and efficiency, and thus bring new breakthroughs and progress in the field of cloud computing resource management.Rational allocation of resources plays a crucial role in cloud computing. In the resource allocation of cloud computing, the cloud computing center has limited cloud resources, and users arrive in sequence. Each user requests the cloud computing center to use a certain number of cloud resources at a specific time.**

## I. INTRODUCTION

In recent years, cloud computing has been widely used. Cloud computing refers to the centralized computing resources, users through the access to the centralized resources to complete the calculation, the cloud computing center will return the results of the program processing to the user. Cloud computing is not only for individual users, but also for enterprise users. By purchasing a cloud server, users do not have to buy a large number of computers, saving computing costs. Cloud service providers can dynamically schedule computing resources according to users' access requirements to maximize computing resource utilization efficiency. According to a report by China Economic News Network, the scale of cloud computing in China has reached 209.1 billion yuan. At present, the more mature cloud service providers in China are Ali Cloud, Baidu Cloud, Huawei Cloud and so on.

Rational allocation of resources plays a crucial role in cloud computing. In the resource allocation of cloud computing, the cloud computing center has limited cloud resources, and users arrive in sequence. Each user requests the cloud computing center to use a certain number of cloud resources at a specific time. The cloud computing service center needs to decide whether to accept the user's request or put the user's request on hold. Reasonable allocation of cloud resources can maximize the quality of service users and improve user satisfaction under limited resources. Most of the existing resource allocation algorithms use heuristic algorithms to allocate resources, but it is difficult for heuristic algorithms to achieve good results in complex situations.

The resource allocation of cloud computing needs to consider various needs of users. Different users have different requirements on cloud resources. The cloud computing center needs to allocate resources according to user requirements. Users' demand for cloud resources can be divided into two types of constraints, namely, hard constraints and soft constraints. Hard constraints refer to the constraints that must be met. For example, a user requests a group of VMS instead of a single VM. To ensure high availability, each VM needs to be placed in a separate fault domain (a server sharing a single point of failure) to avoid economic losses caused by the failure of the cloud computing center. Soft constraints refer to constraints that can or can not be satisfied, but satisfying constraints can significantly improve service quality. For example, in the process of neural network training, placing cloud virtual machines close to each other in the network structure can reduce network latency and speed up computing. During the processing of user requests, VMS that are geographically close to users can improve the communication speed between the cloud computing center and users and improve user experience.

This study mainly proposes a cloud resource allocation method based on deep reinforcement learning considering user needs. This study mainly considers assigning users to servers that are close to improve the service quality of users. However, if users are assigned to servers that are close to them, some servers may be congested and the waiting time will be long. Therefore, this study considers a cloud resource allocation method based on deep reinforcement learning. Deep reinforcement learning can determine the dynamic allocation of resources according to the current state of the system, thus maximizing the utilization efficiency of cloud resources and reducing user waiting time.

## II. RELATED WORK

### A. Resource scheduling by deep reinforcement learning

In their research, Mao et al. (2016) used deep learning methods to explore resource allocation in cloud computing environments. Their research mainly adopts the strategy gradient method to realize the dynamic allocation of resources. The strategy gradient approach is a reinforcement learning technique that maximizes the desired cumulative reward by optimizing the strategy to achieve efficient resource allocation. This approach not only ADAPTS to different environments and requirements, but also enables efficient resource utilization in dynamically changing cloud computing environments. The research of Mao et al. provides important reference and inspiration for the application of deep learning in the field of cloud computing resource management, and provides new ideas and methods for solving resource allocation problems.

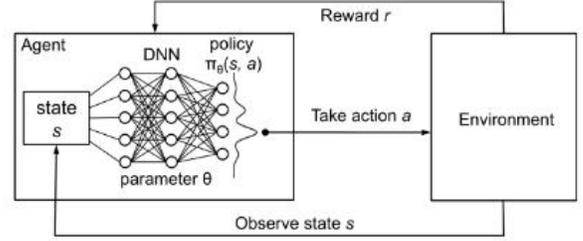

Figure 1. Reinforcement learning methods for policy networks (Figure: Mao et al. (2016))

As shown in Figure 1, the input of the neural network is the current state of the system, and the output is the action. When the system adopts the action, the environment will return the corresponding reward. The goal of system optimization is to minimize the user's waiting time, that is, to maximize it:

$$\nabla_\theta E_{\pi\theta} [\sum_{t=0}^{\infty} \gamma^t r_t] =$$
$$E_{\pi\theta} [\nabla_\theta \log \pi_\theta (s, a) Q^{\pi\theta}(s, a)] \quad (1)$$
$$\nabla_\theta E_{\pi\theta} [\sum_{t=0}^{\infty} \gamma^t r_t] \quad (2)$$

$y \in (0,1)$, where "is the discount factor. The method of strategic gradient descent is mainly to gradient the total reward :-1 where $Q^{\pi\theta}$ (s,a) is the expected cumulative discounted reward for taking action a in state s. The Monte Carlo method is used to calculate the cumulative discount return v, and the parameters of the neural network are updated by the following method.

$$\theta \leftarrow \theta + \alpha \sum_t \nabla_\theta \log \pi_\theta (s_t, a_t) v_t \quad (3)$$

### B. Resource scheduling considering time-varying characteristics

The study of Mondal et al. (2021) focuses on the differences in resource utilization over different time periods when users use cloud computing resources for a long period of time, especially in cloud computing tasks such as neural network training that require a long running time. Considering the different requirements of users for computing resources in different time periods, it can effectively reduce the pressure on servers during peak hours and reduce the risk of server downtime.

In this study, an unsupervised learning method is used to learn the time-varying characteristics of user resource utilization by constructing a learner that utilizes the historical characteristics of user resource use. At the same time, the research puts forward a variety of new reward evaluation mechanisms. The framework of the study is as follows:

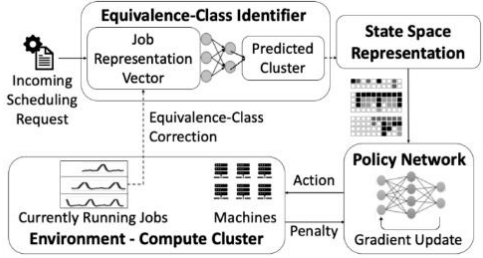

Figure 2. Deep reinforcement learning method considering time-varying features

In this framework, users' historical resource usage data is first collected and analyzed to understand patterns and trends in resource utilization over different time periods. This data is then used to train the model to learn how the user's resource needs change over time. Next, according to the learned pattern, develop a resource allocation strategy to allocate resources reasonably in different time periods, so as to reduce the pressure on the server and reduce the risk of downtime. Finally, a new incentive evaluation mechanism is used to evaluate and optimize the resource allocation strategy to improve the system performance and efficiency. The research of Mondal et al. provides an innovative solution to the resource allocation problem in long-term cloud computing tasks, and makes an important contribution to improving the efficiency and stability of resource utilization in cloud computing environments.

As illustrated in Figure 2, in comparison with section 2.1, this study introduces a novel approach by initially employing clustering techniques to forecast the time-evolving behaviors of users in their utilization of cloud resources. This proactive measure aims to preempt the confluence of peak usage periods among different users, a scenario that would otherwise exert immense strain on the server infrastructure. Within the realm of clustering methodology, a combination of autoregressive analysis, linear trend analysis, and dynamic time warping techniques is predominantly employed. Additionally, the inclusion of dynamic time warping facilitates the identification of users exhibiting similar temporal utilization patterns through the application of K-means clustering. This multifaceted approach not only enhances the predictive capabilities regarding resource utilization dynamics but also lays the groundwork for proactive resource allocation strategies aimed at alleviating server pressure and optimizing resource allocation efficiency.

In the part of deep reinforcement learning, the idea of this study is roughly similar to that of Study 1. Study 2 puts forward a number of definitions of reward worthy of reference. It mainly includes the following parts:

$$P_C = -\sum_d \sum_{m \in M} K_c * C_r(m, d) \quad (4)$$

- Competition

The competition describes the usage of server resources by different resources. If the usage of allocated VMS by two services peaks at the same time, the competition score is higher. Where K is a constant, representing the competitive score in the total prize

The proportion of encouragement (punishment). Cr(m.d) is mainly calculated by the following formula, representing the inner product of the resources used by different services.

$$C_r(m, d) = \sum_{W_i \in m_W} \sum_{W_j \in m_W, j>i} \langle R(W_i, d), R(W_j, d) \rangle \quad (5)$$

- Machine utilization rate

Machine utilization is mainly evaluated for the proportion of machines currently in use. Adding this reward can make reinforcement learning allocate resources using machines rather than non-machine resources in resource allocation.

$$P_U = -\sum_d \sum_{m \in M_u} |U_m(t, d)|^{K_u} \quad (6)$$

Where U (t; d) represents the unused resource of resource d by the used machine m at time t; M. Represents the collection of machines currently in use.

- Excessive use of penalties

$$P_O = -\sum_d \sum_{m \in M} K_o * \mathbb{I}_{m, d}[First\ overshoot\ for\ the\ TVW] \quad (7)$$

Overuse penalties are a measure of penalties given when resource usage is higher than machine usage, where |md is an indicator of overuse.

- Use time penalties

The use of time penalties is mainly calculated by the number of requests waiting in the queue, mainly by the following formula:

$$P_W = -K_\omega * |Q_t| \quad (8)$$

Where $|Q_t|$ represents the number of requests waiting in the queue.

C. Task scheduling/unloading with priority constraints

- Optimization Problem

In the cloud computing resource allocation and task scheduling system, complex tasks are decomposed into multiple subtasks to form task flows and then allocated to processors for parallel processing in order to enhance computational efficiency. As certain tasks' computations depend on the results of previous tasks and priority constraints exist among tasks, directed acyclic graphs (DAGs) can be utilized to abstract and model the workflow. In this model, nodes represent subtasks, and edges between nodes denote priority constraints among subtasks. Literature [4] provides a tree diagram of the DAG scheduling problem, considering factors such as communication time between tasks, processor resource limitations, and interconnectivity between processors.

- Mathematical Model and Related Algorithms

The priority constraints between tasks can be expressed as EST(j) ≤ CT(i), for all i→j, where EST(j) denotes the earliest time task j can begin processing, CT(i) represents the completion time of task i, and i→j indicates that task i must be completed before task j. For the cloud offloading problem on mobile terminals, processors can handle multiple tasks simultaneously. Literature introduces a deterministic delay-constrained task segmentation algorithm based on dynamic programming, demonstrating its suboptimality. Considering the deadline constraints, the topology of DAG is not restricted, assuming processors can only handle one task at a time. Literature [5] presents a mixed-integer programming model considering collaboration between edge computing nodes and remote cloud nodes. This problem involves a general allocation problem with NP-hard characteristics, and there is currently no polynomial-time optimization method. Literature [6] utilizes relaxed integer programming model with 0-1 variables to convert the problem into a convex optimization problem, followed by the design of a heuristic approach. Addressing the multi-DAG mobile terminal offloading problem, literature [7] proposes a mixed-integer programming model to decide whether to upload tasks to the cloud and optimize energy consumption under deadline constraints.

- Heuristic Methods

Optimization methods for DAG scheduling mainly include heuristic methods, intelligent algorithms, and hybrid algorithms, among which heuristic methods are primarily categorized into list scheduling, clustering, and task replication. List scheduling arranges tasks based on priority and selects the highest-priority task from the pending tasks for assignment to the appropriate processor. Clustering groups tasks until the number of categories equals the number of processors. Task replication involves duplicating tasks with significant data to multiple processors to reduce processing latency.

In the "cloud-edge-end" system, literature [6] employs forward ranking as the criterion for list scheduling to optimize the computation, communication costs, and latency of cloud and edge computing nodes. Considering both deadline and cost, literature [7] employs methods such as lower bound estimation to allocate deadlines and compute nodes to DAG subtasks. Referring to the order of task deadlines, literature [8] allocates virtual machines based on the earliest completion time of tasks to optimize overall time performance.

- Intelligent Algorithms

Distinct from heuristic rules, intelligent algorithms aim for global optimization performance. Literature [8] utilizes genetic algorithms to optimize task-edge node group assignments. Probability is employed by literature [9] to characterize the positional relationship between tasks. After DAG pre-segmentation based on heuristic methods, literature [9] utilizes bivariate correlation distribution estimation algorithms to rank tasks and optimize overall application completion time and edge node energy consumption. In literature [10], Estimation of Distribution Algorithm (EDA) is employed to optimize total delay considering task deadline information. For the task-node assignment problem, literature [10] applies particle swarm optimization algorithm to optimize weighted objectives of cost and completion time.

Overall, the above methodologies are logically aligned with the actual requirements of cloud computing resource allocation and task scheduling.

III. EXPERIMENT AND METHODOLOGY

A. Experimental environment

The CloudSim3.0.2 cloud simulation platform of the Grid Laboratory of the University of Melbourne was used in the experiment to test the performance of the author's algorithm. The experiment involved simulation comparison and result analysis with the basic ant colony algorithm (ACO)[10] and simulated annealing algorithm (SA). Firstly, in the cloud simulation platform, the MyAll-ocationTest class is created to perform the initial configuration of the cloud environment. This includes the creation of the data center, initialization of the scale parameters of the cloud computing task, determination of the task size and the size of the input and output data files, and the creation of virtual machine resources. Each VM resource is characterized by the number of CPUs, memory size, bandwidth, and instruction processing speed. Subsequently, cloudsim objects are created to add cloud computing tasks. Moreover, GAACO, ACO, and SA algorithms are implemented in the DatacenterBroker class. The relevant parameters of the genetic ant colony algorithm are presented in Table 1.

TABLE I. GENETIC ANT COLONY ALGORITHM PARAMETER TABLE

| Parameter Symbol | Meaning | Value |
| --- | --- | --- |
| evolution Num | Evolution generations | 100 |
| population | Population size | 10 |
| m | Number of ants | 31 |
| Pc | Crossover probability | 0.35 |
| Pm | Maximum mutation probability | 0.08 |
| A max | Maximum pheromone factor | 1.00 |
| .max | Maximum expected pheromone factor | 2.00 |
| Y max | Maximum pheromone evaporation coefficient | 0.10 |
| Q | Maximum pheromone intensity | 50.00 |

The experimental design and result analysis in this paper focus on comparing the performance of the genetic ant colony algorithm across four key aspects: average time cost, average cost, algorithm service quality, and system resource load rate.

*B. Experimental parameter*

Initially, the task size is set to 10, with cloud computing resources consisting of 10 VMs. Each VM has a storage size of 10 GB, memory size of 256 MB, one CPU, and a bandwidth of 1,000 MB. The unit time bandwidth cost and unit time instruction cost are 0.01 yuan/s each. Tasks are experimented with in increments of 10.

The quality of service for the algorithm is represented by multiQoS. The resource load rate is defined as follows:

$$\text{resource load rate} = \frac{\text{use}_i}{\text{useAvg} \times n} \quad (9)$$

The average time cost for each algorithm is calculated as the number of tasks increases by 10. The results are depicted in Figure 1. It's observed that the time cost of GAACO is superior to that of ACO, albeit longer than SA. Moreover, as the number of tasks increases, the time gap widens, with GAACO reducing time by 50.9% compared to ACO and showing a 3% difference compared to SA. Thus, the author's algorithm outperforms ACO in terms of time cost, albeit with a slight difference compared to SA.

The cost of each algorithm under the condition of different number of tasks is shown in Figure 3. The number of tasks is increased by a multiple of 10. It can be seen that the difference between algorithms is not large, and the average cost is only about 1%.

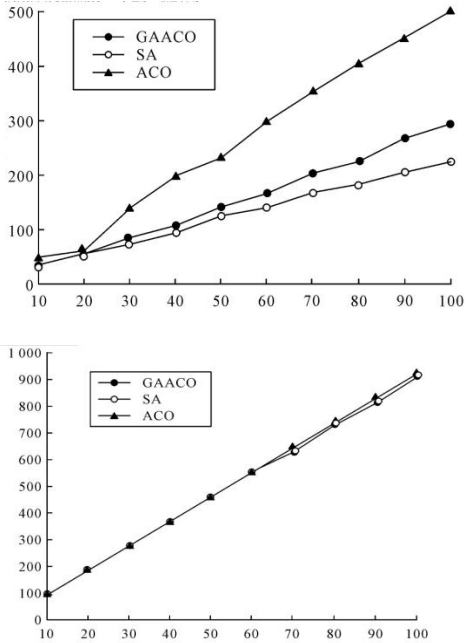

Figure 3. Average time cost of each algorithm and Cost of each algorithm

The experimental results of service quality of each algorithm are shown in Figure 5. It can be seen that the service quality of the author's algorithm and SA increases slowly with the increase of the number of tasks, while ACO presents a linear and sharp rise. Service quality is a comprehensive index of cost, time and reliability, and it can be seen that the comprehensive performance of GAACO is better than that of ACO and SA. They reduced by 14.4% and 76.8%, respectively.

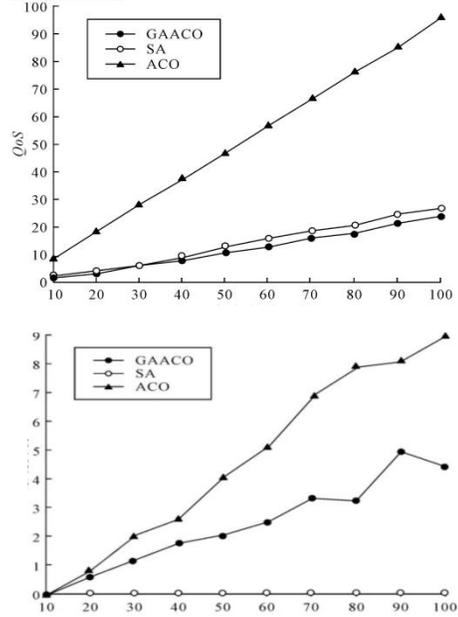

Figure 4. Service quality of each algorithm and system load of each algorithm

The experimental results regarding the algorithm's system load are presented in Figure 4. It's evident that the system load of ACO has consistently remained high, whereas GAACO exhibits a higher load compared to SA but is notably superior to ACO. Specifically, GAACO achieves a 50.2% reduction in average system load compared to ACO.

Furthermore, in conjunction with Figures 2 and 3, it's observed that SA tends towards an evenly distributed task assignment to virtual machines, resulting in a system load of 0 as per equation (9). However, effective cloud computing task scheduling necessitates a balance between cost, time, and reliability, as reflected in QoS. Therefore, the author's proposed algorithm better aligns with customer requirements during the actual scheduling process.

*C. Experimental conclusion*

After conducting thorough research on cloud computing task scheduling, the proposed algorithm has been rigorously compared with both the basic ant colony algorithm and the simulated annealing algorithm across four critical aspects. The comprehensive analysis of the results reveals that the proposed algorithm consistently outperforms the other two algorithms. Notably, the proposed algorithm demonstrates superior capabilities in balancing various factors including time cost, monetary cost, reliability, and system load. This balanced optimization ensures that the algorithm can effectively meet the multidimensional quality of service (QoS) requirements of users.

In essence, the experimental findings underscore the efficacy of the proposed algorithm in addressing the complex challenges inherent in cloud computing task scheduling. By surpassing traditional methods, it offers a more holistic and efficient approach to meeting the diverse needs of users while ensuring optimal resource utilization and performance. Thus, the proposed algorithm stands as a promising solution for enhancing cloud computing task scheduling processes in practical applications.

## IV. REALIZE DYNAMIC ALLOCATION OF RESOURCES

Dynamic allocation of resources is a crucial aspect of resource management in cloud computing environments. To achieve this, several key components and processes need to be considered:

### A. Resource Scheduling Algorithm

The resource scheduling algorithm serves as the foundation for dynamic resource allocation. It evaluates various factors such as resource demand, type, and availability to select the optimal scheduling scheme. Common algorithms include load balancing, static allocation, and dynamic allocation. Each algorithm addresses specific requirements and constraints to ensure efficient resource utilization.

### B. Resource Monitoring

Resource monitoring is essential for detecting changes and anomalies in real-time. Different types of resources must be continuously monitored to track usage patterns and identify potential issues. By monitoring resources, organizations can proactively address fluctuations in demand and optimize resource allocation strategies accordingly.

### C. Resource Forecasting

Forecasting future resource demand and usage trends is critical for effective resource allocation. Historical data analysis and prediction techniques are used to anticipate future resource requirements. By accurately forecasting resource needs, organizations can preemptively allocate resources to prevent shortages and optimize utilization, minimizing waste and maximizing efficiency.

### D. Resource Management

Effective resource management is essential for facilitating dynamic resource allocation. Resources must be properly classified, labeled, and organized to enable efficient location and scheduling. Through robust resource management practices, organizations can streamline operations, improve resource utilization, and optimize performance.

By integrating these components into a cohesive framework, organizations can realize the full potential of dynamic resource allocation in cloud computing environments. This approach ensures that resources are allocated efficiently, adaptively, and in accordance with evolving demands and conditions, ultimately enhancing productivity and driving business success.

## V. CONCLUSIONS

In conclusion, this study presents a comprehensive approach to address the challenges of resource scheduling and management in cloud computing environments. By leveraging machine learning optimization techniques, particularly deep reinforcement learning, the proposed algorithm demonstrates significant improvements in system performance and efficiency. Through extensive experimentation and analysis, it is evident that the proposed algorithm outperforms traditional methods such as ant colony optimization and simulated annealing in terms of time cost, cost effectiveness, service quality, and system resource load. This underscores its potential to revolutionize cloud computing resource management and bring about new breakthroughs in the field.

### A. Synergy of Deep Learning and Cloud Computing Scheduling

The integration of deep learning techniques with cloud computing scheduling presents a promising avenue for achieving more intelligent and adaptive resource allocation strategies. Deep reinforcement learning, as demonstrated in this study, enables dynamic allocation of resources based on the current system state, leading to maximized resource utilization efficiency and reduced user waiting times. By considering user needs and system constraints in a holistic manner, deep learning-based approaches can offer personalized and optimized solutions to complex scheduling problems in cloud environments.

### B. Advantages of Deep Learning in Cloud Computing Scheduling

Deep learning brings several advantages to cloud computing scheduling. Firstly, its ability to learn complex patterns and relationships from data enables it to adapt to diverse and evolving cloud environments. Secondly, deep learning models can handle high-dimensional and non-linear data, allowing for more accurate and nuanced decision-making. Additionally, deep learning algorithms can continuously improve over time through experience, leading to enhanced performance and scalability in cloud scheduling tasks. Overall, the integration of deep learning with cloud computing scheduling holds great promise for addressing the increasingly complex and dynamic nature of modern cloud environments.

### C. Future Prospects of AI and Cloud Computing Integration

Looking ahead, the convergence of artificial intelligence (AI) and cloud computing is expected to drive significant advancements in various domains. AI-powered cloud services will offer more intelligent and autonomous capabilities, such as predictive resource provisioning, anomaly detection, and self-optimizing systems. Furthermore, AI-driven insights derived from vast amounts of cloud data will enable businesses to make more informed decisions and gain competitive advantages. As AI technologies continue to evolve, their integration with cloud computing will usher in a new era of innovation and transformation across industries, paving the way for smarter, more efficient, and more resilient digital ecosystems.